\documentclass[10pt,journal]{IEEEtran}

\usepackage{amsmath,times,amssymb,epsfig,nicefrac,color,balance}
\usepackage{amsfonts,graphics,subfigure,epsfig,graphics}
\usepackage[mathcal]{euscript} 

\usepackage{mathrsfs}
\usepackage{cite,calc}
\usepackage{euler}
\newcommand\nc\newcommand
\nc\bfa{{\boldsymbol a}}\nc\bfA{{\boldsymbol A}}\nc\cA{{\mathcal A}}
\nc\bfb{{\boldsymbol b}}\nc\bfB{{\boldsymbol B}}\nc\cB{{\mathcal B}}
\nc\bfc{{\boldsymbol c}}\nc\bfC{{\boldsymbol C}}\nc\cC{{\mathcal C}}
\nc\sC{{\mathscr C}}
\nc\bfd{{\boldsymbol d}}\nc\bfD{{\boldsymbol D}}\nc\cD{{\mathcal D}}
\nc\bfe{{\boldsymbol e}}\nc\bfE{{\boldsymbol E}}\nc\cE{{\mathcal E}}
\nc\bff{{\boldsymbol f}}\nc\bfF{{\boldsymbol F}}\nc\cF{{\mathcal F}}
\nc\bfg{{\boldsymbol g}}\nc\bfG{{\boldsymbol G}}\nc\cG{{\mathcal G}}
\nc\bfh{{\boldsymbol h}}\nc\bfH{{\boldsymbol H}}\nc\cH{{\mathcal H}}
\nc\bfi{{\boldsymbol i}}\nc\bfI{{\boldsymbol I}}\nc\cI{{\mathcal I}}
\nc\bfj{{\boldsymbol j}}\nc\bfJ{{\boldsymbol J}}\nc\cJ{{\mathcal J}}
\nc\bfk{{\boldsymbol k}}\nc\bfK{{\boldsymbol K}}\nc\cK{{\mathcal K}}
\nc\bfl{{\boldsymbol l}}\nc\bfL{{\boldsymbol L}}\nc\cL{{\mathcal L}}
\nc\bfm{{\boldsymbol m}}\nc\bfM{{\boldsymbol M}}\nc\sM{{\mathscr M}}
\nc\bfn{{\boldsymbol n}}\nc\bfN{{\boldsymbol N}}\nc\cN{{\mathcal N}}
\nc\bfo{{\boldsymbol o}}\nc\bfO{{\boldsymbol O}}\nc\cO{{\mathcal O}}
\nc\bfp{{\boldsymbol p}}\nc\bfP{{\boldsymbol P}}\nc\cP{{\mathcal P}}
\nc\bfq{{\boldsymbol q}}\nc\bfQ{{\boldsymbol Q}}\nc\cQ{{\mathcal Q}}
\nc\bfr{{\boldsymbol r}}\nc\bfR{{\boldsymbol R}}\nc\cR{{\mathcal R}}
\nc\bfs{{\boldsymbol s}}\nc\bfS{{\boldsymbol S}}\nc\cS{{\mathcal S}}
\nc\bft{{\boldsymbol t}}\nc\bfT{{\boldsymbol T}}\nc\cT{{\mathcal T}}
\nc\bfu{{\boldsymbol u}}\nc\bfU{{\boldsymbol U}}\nc\cU{{\mathcal U}}
\nc\bfv{{\boldsymbol v}}\nc\bfV{{\boldsymbol V}}\nc\cV{{\mathcal V}}
\nc\bfw{{\boldsymbol w}}\nc\bfW{{\boldsymbol W}}\nc\cW{{\mathcal W}}
\nc\bfx{{\boldsymbol x}}\nc\bfX{{\boldsymbol X}}\nc\cX{{\mathcal X}}
\nc\bfy{{\boldsymbol y}}\nc\bfY{{\boldsymbol Y}}\nc\cY{{\mathcal Y}}
\nc\bfz{{\boldsymbol z}}\nc\bfZ{{\boldsymbol Z}}\nc\cZ{{\mathcal Z}}

\DeclareMathOperator{\rank}{rank}

\newcommand{\remove}[1]{}

\newcommand{\dist}{d_\mathrm{H}}
\newcommand{\rdss}{\mathbb{CAP}}
\newcommand{\indx}{\mathrm{INDEX}}
\newcommand{\len}{\mathrm{len}}
\newcommand{\minrank}{\mathrm{minrank}}

\newtheorem{theorem}{Theorem}
\newtheorem{definition}{Definition}
\newtheorem{lemma}[theorem]{Lemma}
\newtheorem{proposition}[theorem]{Proposition}

\newtheorem{remark}{\indent Remark}
\newtheorem{example}{Example}

\newcommand\ff{{\mathbb F}}

\newcommand\integers{{\mathbb Z}}
\newcommand\rationals{{\mathbb Q}}

\begin{document}
\sloppy

\title{Storage Capacity of Repairable Networks}
\author{Arya Mazumdar~\IEEEmembership{Member,~IEEE}
\thanks{The author is with the Department of Electrical and Computer Engineering, University of Minnesota, Minneapolis, MN  55455, email: \texttt{arya@umn.edu}.
  Part of this work was presented in the IEEE International Symposium on Information Theory, 2014 \cite{mazumdar2013duality} and in 
   the Allerton Conference, 2014 \cite{mazumdar2014achievable}.
  This work was supported in part by the National Science Foundation CAREER award under grant no. CCF 1453121.}}

\allowdisplaybreaks
\maketitle

\begin{abstract}
In this paper,  we introduce a model of a distributed storage system 
that is  locally recoverable from any single server failure. Unlike the usual local recovery model of
codes for distributed storage,
this model accounts for the fact that 
each server or storage node in a network is 
connectible to only some, and not all other, nodes.
 This may happen for reasons such as  physical separation, inhomogeneity 
in storage platforms etc.
We estimate the storage capacity of both undirected and directed networks under this model and propose some
constructive schemes. From a coding theory point of view, we show that this model
is  {\em approximately}
dual of the well-studied index coding problem. 

Further in this paper, we extend the above model to handle multiple server failures. Among other results, we provide an upper bound
on the minimum pairwise distance of a set of words that can be stored in a graph with the local repair 
guarantee.  The well-known impossibility bounds on the distance of locally recoverable codes  follow from our result.
\end{abstract}

\section{Introduction}

Recently, the  local repair property of error-correcting codes is the center of a lot of research activities. 
In a distributed storage system, a single server  failure is the most common error-event, and in the 
case of a failure the aim is to reconstruct the content of the failed server from as few other servers as possible (or by downloading 
minimal amount of data from other servers). The study of such {\em regenerative}  storage systems was
initiated in \cite{dimakis2010network} and then followed up in several recent works. In \cite{gopalan2012locality},
a particularly neat characterization of the local repair property is provided. It is assumed that,  each symbol of an encoded 
message is stored at a different
node in the storage-network (since the symbol alphabet is
  unconstrained, a symbol could represent a packet or block of bits of
  arbitrary size). 
  Accordingly,
\cite{gopalan2012locality} investigates code-families that allow any
single coordinate of a codeword to be recovered from at most a constant
number of other coordinates of the codeword, i.e., from a number of
coordinates that does not grow with the length of the code. 

The work of \cite{gopalan2012locality} is then further generalized to several directions
and a number of impossibility results and constructions of {\em locally
repairable codes} were presented  in 
 \cite{papailiopoulos2012locally,tamo2013optimal,cadambe2013upper,silberstein2013optimal,kamath2012codes, barg2013family}
 among others. The central result of this body of works is that for any code of length $n$,
 dimension $k$ and minimum distance $d$, 
 \begin{equation}\label{eq:locality}
 d \le n -k -\Big\lceil\frac{k}{r}\Big\rceil +2,
 \end{equation}
 where $r$ is such that any
single coordinate can be recovered from at most $r$ other coordinates \cite{gopalan2012locality}.

However, the topology of the network of
distributed storage system is missing from the above definition of local repairability.
Namely, all servers are treated equally irrespective of their physical positions, proximities, 
and connections.
Here, in this paper, we take a step to include the network topology into consideration.
 We study the case when the architecture of the storage system is fixed and the network of storage
 is given by a graph. In our model, the servers are represented by the vertices of a graph, and
 two servers are connected by an edge if it is easier to establish up-or-down link between them, for reasons
 such as physical locations  of the servers, architecture of the distributed system or 
 homogeneity of softwares, etc. It is reasonable to assume that
 the storage-graph is directed, because there may be varying difficulties in establishing
 an up or down link between two servers.
 Under this model, we impose the local recovery or repair condition in the following way:
 the content of any failed server must be reconstructible from the neighboring servers on {\em the storage graph}.
 
  Assuming the above model, the main quantity of interest is the amount of information that
 can be stored in the graph. We call this quantity the {\em storage  capacity} of the graph. Finding this
 capacity exactly, as well as to construct explicit schemes that achieve this capacity, are both
 computationally hard problems for an arbitrary graph. However, we show that good approximation 
 schemes are possible -- and for some special classes of graphs we can even compute this capacity 
 exactly with constructive schemes. In particular, for any undirected graph, the storage capacity is
 sandwiched between the {\em maximum matching} and the {\em minimum vertex cover}, two quantities
 within a factor of two of each other. Similar 
 statement, albeit concerning different properties, is possible for directed graphs.
 
 It turns out that,
our model is closely related to the popular {\em index coding} problem on a side information graph.
In the index coding problem, a set of users are assigned bijectively to a set of variable that they want to know.
However, instead of knowing the assigned variable, each knows a subset of other variables. 
This scenario can be depicted by a so-called {\em side-information} graph where each vertex represents a user and
 there is an edge between users A and B, if A knows the variable assigned to B. Given this graph, how much
 information should a broadcaster has to transmit, such that each vertex can deduce its assigned variable? 

The above problem of index coding was introduced in
\cite{bar2006index} (it has a predecessor in \cite{birk2006coding}), and since then is a subject of extensive research. 
It was shown in \cite{el2010index} that any network coding problem can be reduced to an index coding problem,
and the index coding capacity is among the computationally hardest problems of all network coding \cite{langberg2008hardness,blasiak2011lexicographic}.
A prominent work in the index coding literature is \cite{alon2008broadcasting}, that studies the {\em broadcast rate} for index coding.
It turns out that an auxiliary  quantity (called $\gamma$) used in \cite{alon2008broadcasting} is exactly the storage capacity\footnote{Actually, $\log \gamma$ is the storage capacity that we introduce here.}
that we introduce and study in this paper (attachment of $\gamma$ to any quantity of practical use  was absent in \cite{alon2008broadcasting}). Recently K. Shanmugam  \cite{shanmu} pointed out to the author that this quantity has also been studied as {\em graph entropy}\footnote{In literature, the term ``graph entropy'' usually refers to a different quantity \cite{korner1973coding}.} in \cite{riis2007graph} by Riis. 

Using the results of \cite{alon2008broadcasting} it is possible to
show a connection between the broadcast rate of index coding and the storage-capacity when the side-information graph
and the storage graph are the same. Indeed, we show that there exists a {\em duality}
 between a storage code and an index code on the same graph. This observation, which also connects the
 {\em complementary index coding} rate of \cite{chaudhry2011complementary} with the storage capacity, is further explored in this paper.  

The local repairability property on a graphical model of storage can be extended to several directions. 
One may ask for protection against catastrophic failures, and therefore also impose a minimum {\em distance}
condition on codes, which is a common fixture of the local recovery literature. 
In this scenario, we obtain a general bound that include previous results such as eq.~\eqref{eq:locality} as special
cases. Moreover such bounds can also be made dependent on the size of the alphabet (size of storage node).

Furthermore, instead of a single
node local repairability, multiple failures can also be considered.
Such multiple failures and the corresponding cooperative local recovery model in distributed storage 
was recently introduced in \cite{rawat2014cooperative}. In this paper we generalize this model on graphs.

The storage coding problem of our model is  a very fundamental network coding problem, and one of our main observation is that
reasonable approximation schemes are possible for storage coding. While the index coding rate is very hard to approximate (see, \cite{langberg2008hardness})
it is possible to have good approximation constructively for storage capacity with {\em linear} (explained in Section~\ref{sec:rdss}) codes.
This should be put into contrast with results, such as \cite[Thm.~1.2]{blasiak2011lexicographic}, which show that 
a rather large gap must exist between vector linear and nonlinear index coding (or general network coding) rates.

Apart from the approximation guarantee, there are other evidences to the fact that index coding and our storage coding
 are two very different problems by nature. For example, for two disconnected graphs, the total storage capacity is the
 sum of the
 capacities of the individual graphs. But the index coding length for the union of two disconnected graphs may be smaller than the
 sum of individual code lengths of the graphs  (see, Thm.~1.1-1.4 and the accompanying discussions in \cite{alon2008broadcasting}).

  In  a parallel independent work  \cite{shanmugam2014bounding},
one of our initial results, namely,
 the duality between storage and index codes (see Prop.~\ref{prop:dual_lin}) is proved  for {\em  vector linear codes}. The authors of  \cite{shanmugam2014bounding} further
 use that observation
 to  give an upper bound on the optimal
linear sum rate of the multiple unicast network coding problem. In this paper we have a completely different focus. 

\subsection{Results and organization}

The paper is organized in the following way. 

\begin{itemize}
\item \noindent{\em Model of a repairable distributed storage:}
In Section \ref{sec:rdss}, we introduce formally
the model of a recoverable distributed storage system and the notion
of an optimal 
storage code given a graph.
This section also introduces the quantities of our interest, namely the capacity of storage.

\item \noindent{\em Relation to index coding:}
In Section \ref{sec:index}, we 
 explore  the relation  of an optimal storage code  to the optimal 
index code. We provide an algorithmic
proof of a duality relation between the index code and distributed storage code. Our proof is
based on a covering argument of the Hamming space, and rely on the fact that for any given subset of the Hamming space
there exist several translations of the set, 
  that have very small overlaps with the original subset. 
  
\item \noindent{\em Bounds and algorithms:}  
  In Section~\ref{sec:undirected}, we 
  provide constructive schemes that achieves a storage-rate within
  half of what is maximum possible for any undirected graph (the scheme is optimum for bipartite graphs). Some other existential
 results are also proved in this section. Next, we extend the approximation schemes 
 towards directed graphs in Section~\ref{sec:directed}. It turns out to be a  harder problem for directed graphs
 and we provide
 a scheme with
 a logarithmically (with graph size) growing approximation factor. 
 
\item \noindent{\em Bounds on minimum distance and other multiple failure models:}  
 In the last section, Section~\ref{sec:gen}, we generalize the notions of local recovery on graphs to 
 include the minimum distance criterion and cooperative local recovery.  In both of these cases, we provide
  fundamental converse bounds and outline some constructive schemes. In particular, the well-known impossibility results
    on the
  minimum distance  of a locally repairable codes, 
  such as eq.~\eqref{eq:locality} 
  or the ones presented in 
 \cite{cadambe2013upper},
 simply follow from Thm.~\ref{thm:bound} and Prop.~\ref{prop:alpha}.
%
%
\end{itemize}

\section{Recoverable distributed storage systems}\label{sec:rdss}
\begin{figure}[t]
\begin{center}
\includegraphics[width=0.5\textwidth]{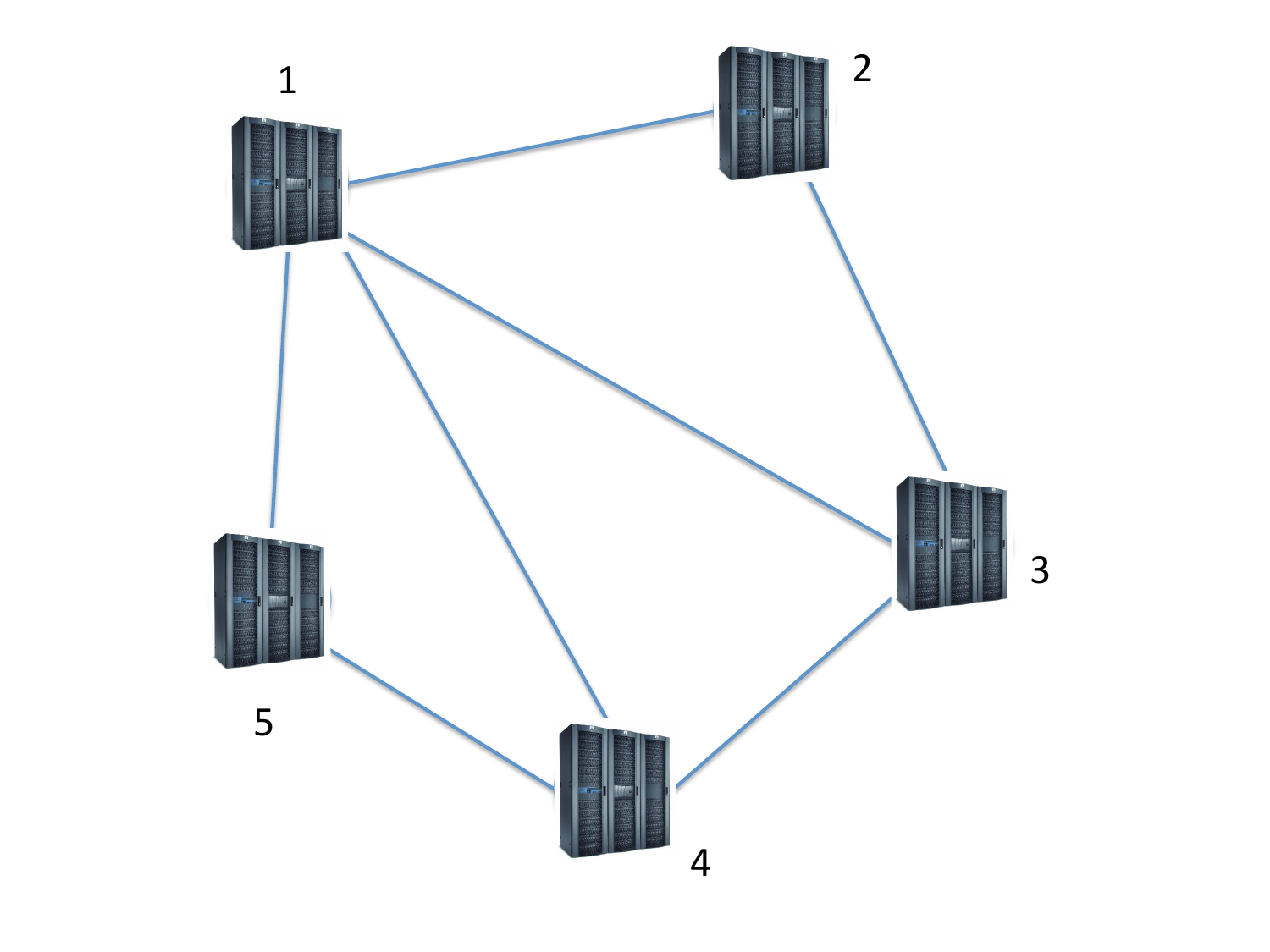}
\end{center}
\caption{Example of a distributed storage graph} 
\label{fig:example}
\end{figure}

In this section, we introduce the basic notion of a single-failure recoverable storage system.
Consider a network of distributed storage, for example, one  of Fig.~\ref{fig:example}, where
several servers (vertices) are connected via network links (edges). As mentioned in the introduction,
the property of
two servers  connected by an edge is based on the ease of establishing a link between the servers\footnote{One might consider a nonnegative weight on 
each edge, which would be a natural generalization}.
If the data of any one server is lost, we want to recover it from the {\em nearby} servers, i.e., the
ones with whom it is easy to establish  links. This notion is formalized below.
It is also possible (and sensible, perhaps) to model this as a directed graph (especially when uplink and downlink 
constructions have varying difficulties). In the rest of the paper the definitions, claims and arguments hold
for both directed and undirected graphs, unless otherwise specified.



Suppose, the  graph $G(V,E)$ represents the network of storage. 
For any $v\in V$, define $N(v)= \{u\in V: (v,u) \in E\}$ to be 
the neighborhood of $v$. 
Each element of $V$
represents a server, and in the case of a server
failure (say, $v \in V$ is the failed server) one must be able to reconstruct its content from its
neighborhood $N(v)$. 

Given this constraint  what is the maximum amount of information one can store in the system?
Without loss of generality, assume $V= \{1,2,\dots,n\}$ and the  variables
 $X_1, X_2,\dots,X_n$ respectively denote the content of the vertices, where, $X_i \in \ff_q, i =1,\dots,n.$ Also for any $I \subseteq V$, let $X_{I} \in \ff_q^{|I|}$ be the projection of  $(X_1, X_2,\dots,X_n)^T$ on to the coordinates of $I$.
 \begin{definition}\label{def:rdss}
A  {\em recoverable distributed storage system (RDSS)
code} $\cC \subseteq \ff_q^n$ with storage recovery graph $G(V,E), V = \{1,2,\dots,n\},$ is a set of
vectors in $\ff_q^{n}$ together with
a set of deterministic recovery  functions, $f_i:\ff_q^{|N(i)|}\to \ff_q$ for $i = 1,\dots,n$ such that
for any codeword $(X_1, X_2,\dots,X_n)^T \in \ff_q^n,$
\begin{equation}
X_i = f_i(X_{N(i)}), \quad i = 1,\dots,n.
\end{equation}
The  decoding functions
 depend on G. The log-size of the code, $\log_q |\cC|$, is called  the dimension  of $\cC$, or $\dim(\cC)$. 
Given a graph $G$ the maximum possible dimension of an RDSS code is denoted by $\rdss_q(G)$.
\end{definition} 
Note that, in this paper, $\rdss_q(G)$ is expressed in $q$-ary units. To convert it to {\em bits} we need to multiply 
with $\log_2 q$. 
 
 As an example, if $G$ is a complete graph then $\rdss_q(G) = n-1$. This is possible because in $n-1$ vertices we can store arbitrary 
 values, and in the last vertex we can store the sum (modulo $q$) of the stored values.
 
%

As another example, consider the graph of Fig.~\ref{fig:example} again. Here, $V = \{1,2,3,4,5\}$. The recovery sets
of each vertex (or storage nodes) are given by:
\begin{align*}
N(1) = \{2,3,4,5\}, \, N(2)  = \{1,3\}, N(3)  = \{1,2,4\}, \\
 N(4)  = \{1,3,5\},  N(5)  = \{1,4\}.
\end{align*}
Suppose, the contents of the nodes $1,2,\dots,5$ are $X_1, X_2,\dots,X_5$ respectively, where, $X_i \in \ff_q, i =1,\dots,5.$
Moreover, $X_1 = f_1(X_2,X_3,X_4,X_5), X_2 = f_2(X_1,X_3), X_3 = f_3(X_1,X_2,X_4), X_4 = f_4(X_1,X_3, X_5), X_5 = f_5(X_1, X_4).$

Assume, the functions $f_i, i =1, \dots, 5$, in this example are linear. That is, for $\alpha_{ij} \in \ff_q, 1\le i,j\le 5,$
\begin{align*}
X_1 &= \alpha_{12} X_2 +\alpha_{13} X_3 + \alpha_{14}X_4 + \alpha_{15}X_5\\
X_2 &= \alpha_{21}X_1+\alpha_{23}X_3\\
X_3 &= \alpha_{31}X_1 +\alpha_{32}X_2+\alpha_{34}X_4\\
X_4 &= \alpha_{41}X_1 + \alpha_{43}X_3 + \alpha_{45}X_5\\
X_5 &= \alpha_{51}X_1+\alpha_{54}X_4.
\end{align*}
This implies, $(X_1,X_2,\dots, X_5)$ must belong to the null-space (over $\ff_q$) of 
\[
D \equiv
 \left( \begin{array}{ccccc}
1 & -\alpha_{12} & -\alpha_{13}  & - \alpha_{14} & -\alpha_{15} \\
-\alpha_{21} & 1 &  -\alpha_{23} & 0 & 0 \\
-\alpha_{31} & -\alpha_{32} & 1 & -\alpha_{34} & 0\\
-\alpha_{41} & 0 &-\alpha_{43}  & 1 & -\alpha_{45}\\
-\alpha_{51} &0 & 0 & -\alpha_{54} & 1
\end{array} \right).\] 
The dimension of the null-space of $D$ is $n$ minus the rank of $D$. At this point the following definition is useful.

Suppose, $A = (a_{ij})$ be an $n \times n$ matrix over $\ff_q$. It is said that $A$ {\em fits}
$G(V,E)$ over $\ff_q$ if $a_{ii} \ne 0$ for all $i$ and $a_{ij} =0$ whenever $(i,j) \notin E$ and $i \ne j$. 
\begin{definition}\label{def:minrk}
The minrank \cite{haemers1978} of a graph $G(V,E)$ over $\ff_q$ is defined to be,
\begin{equation}
\minrank_q(G) = \min\{\rank_{\ff_q}(A): A \text{ fits } G\}.
\end{equation}
\end{definition}   

Notice, in the example above, $D$ {\em fits} the graph $G$. Hence, it is evident
that the dimension of the RDSS code is $n -\minrank_q(G)$ (see, Defn.~\ref{def:minrk}). 
From the above discussion, we have, 
\begin{equation}
\rdss_q(G) \ge n - \minrank_q(G),
\label{eq:minrk}
\end{equation}
and, $n -\minrank_q(G)$ is the maximum possible dimension of an RDSS code when
the recovery functions are all linear. 

Linear RDSS codes are not optimal all the time. This is shown in the following example.

\begin{figure}[t]
\begin{center}
\includegraphics[width=0.5\textwidth]{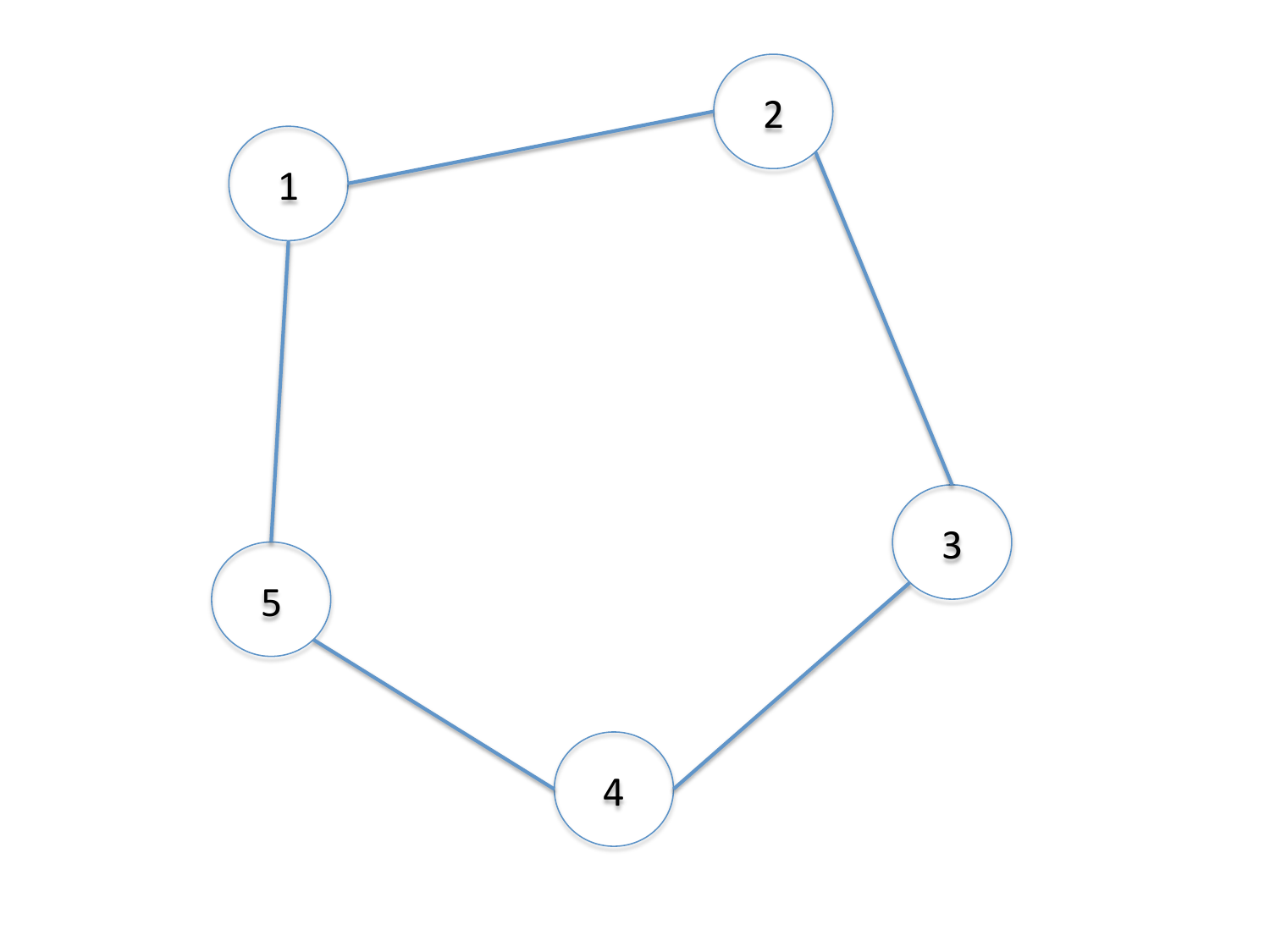}
\end{center}
\caption{A distributed storage graph (the pentagon) that shows $\rdss(G) \ne n -\indx(G)$.} 
\label{fig:pent}
\end{figure}

\begin{example}\label{exm:1}
This example is present in \cite{alon2008broadcasting}, and the distributed storage graph, a {\em pentagon}, is 
shown in Fig.~\ref{fig:pent}. For this graph, a maximum-sized binary RDSS code consists of the codewords
$\{00000,01100,00011,11011,11101\}$. The recovery functions are given by,
\begin{align*}
X_1 = X_2 \wedge X_5, 
X_2 = X_1 \vee X_3,
X_3 = X_2 \wedge \bar{X}_4, \\
X_4 = \bar{X}_3 \wedge X_5,
X_5 = X_1\vee X_4.
\end{align*}
 Here $\rdss_2(G) = \log_2 5$ bits. 
If all the recovery functions are linear, we could not have an RDSS code with 
so many codewords. Indeed, since minrank of this graph over $\ff_2$ is 3, we could have had only 
$2^{5-3}=4$ codewords with linear recovery functions.
\end{example}

Literatures of distributed storage often considers {\em vector  codes} and {\em vector linear codes}.
In our case, in  a vector code, instead of a symbol, a vector is stored in each of the vertices. 
In the context of general nonlinear
codes, vector codes do not bring any further technical novelty and can just be thought of as
codes over a larger alphabet. The capacity of storage can only increase when we 
consider codes over larger alphabet. 
\begin{definition}
Define the {\em vector capacity} of a graph $G(V,E)$ to be,
\begin{equation}\label{eq:vector}
\rdss(G) = \lim_{p \to \infty}  \rdss_{2^p}(G) . 
\end{equation}
\end{definition}
Recall that $\rdss_q(G)$ is measured in $q$-ary units above.
The limit of \eqref{eq:vector} exists and is equal to $\sup_p \rdss_{2^p}(G).$  
This follows from the
superadditivity, $$(p_1+p_2)\rdss_{2^{p_1+p_2}}(G) \ge p_1\rdss_{2^{p_1}}(G)+  p_2\rdss_{2^{p_2}}(G),$$ and Fekete's lemma.

A  vector {\em linear} RDSS
code,  on the other hand, is quite different from  simple linear codes. Each server stores a vector of $p$ symbols, say.
Now in the event of a node failure, each of the $p$ lost symbols must be recoverable by a {\em linear function}
of all the symbols stored in the neighboring vertices. In other words, if $q$-ary symbols are stored in the vertices, then the recovery functions
are over $\ff_q$, and not over $\ff_{q^p}$ (for nonlinear recovery, this does not make any difference).


 The Shannon capacity \cite{shannon1956zero} of a graph is a well-known quantity and
it is known to be upper bounded by minrank \cite{haemers1978}. 
Any concrete reasoning relating  Shannon capacity to $\rdss$ is of interest, but has not been pursued in this paper.
 It is to be noted that for the pentagon of Fig.~\ref{fig:pent},
the Shannon capacity is $\sqrt{5}$ while $\rdss_2(G)= \log_2 5$.

\remove{During some intermediate steps of proofs in this paper, we refrained from using ceiling and floor functions for clarity.
In many cases, it is clear that the number of interest is not an integer and should be rounded
off to the nearest larger or smaller integer. The main results do not change 
for this.  }

\section{Relation with Index Coding}\label{sec:index}
We start this section with the definition of an index coding problem. The main objective of this section
is to establish and explore the relation of the index coding rate and $\rdss_q(G)$, given a graph $G$.

  In the index coding problem, a possibly {\em directed} { side information} graph
$G(V,E)$ is given. Each vertex $v \in V$ represents a receiver that is interested in knowing a uniform random variable $Y_v \in \ff_q$.
 The receiver at $v$ knows the values of the  variables $Y_u, u \in N(v)$.
How much information should a broadcaster  transmit, such that every receiver knows
the value of its desired random variable? Let us give the formal definition from \cite{bar2006index},
adapted for $q$-ary alphabet here.
\begin{definition}
An  {\em index
code} $\cC$ for $\ff_q^n$ with side information graph $G(V,E), V = \{1,2,\dots,n\},$ is a set of
codewords in $\ff_q^{\ell}$ together with:
\begin{enumerate}
\item An encoding function $f$ mapping inputs in $\ff_q^n$
to codewords, and
\item A set of deterministic decoding functions $g_1,\dots,g_n$ such
that $g_i\Big(f(Y_1,\dots,Y_n), Y_{N(i)}\Big) = Y_i$ for every $i=1, \dots,n$.
\end{enumerate}
 The encoding and decoding functions 
 depend on G. The integer $\ell$ is called  the length of $\cC$, or $\len(\cC)$. 
Given a graph $G$ the minimum possible length of an index code is denoted by $\indx_q(G)$.
\end{definition} 

It is not very difficult to deduce the connection between the length of an index code to 
the minrank of the graph -- and it was shown in \cite{bar2006index} that,
\begin{equation}
\indx_q(G) \le \minrank_q(G).
\label{eq:bary}
\end{equation} 
The above inequality can be strict in many cases \cite{alon2008broadcasting,lubetzky2009nonlinear}.
However, $\minrank_q(G)$ is the minimum length of an index code on $G$ when the encoding function, and the decoding functions
are all {\em linear}. The following proposition is also immediate.
\begin{proposition}\label{prop:dual_lin}
 The null-space of a linear index code for $G$ is a linear RDSS
code for the same graph $G$.
\end{proposition}
\begin{IEEEproof}
All the vectors that are mapped to zero by the encoding function of an index code form an RDSS code,
as any symbol stored at a vertex can be recovered by the corresponding index coding decoding function for that vertex.
On the other hand the cosets of an RDSS code partition the space and the set of cosets is isomorphic to the 
null-space of the RDSS code. Hence an index code can be formed that  encodes a vector to the coset it belongs to.
\end{IEEEproof}

Note that, it is not true that
$\rdss_q(G) = n - \indx_q(G)$, although Eq.~\eqref{eq:bary} and Eq.~\eqref{eq:minrk} suggest a similar relation. This is shown 
in the graph of  Example~\ref{exm:1}. There,  the minimum length of
an index code for this graph is $3$, i.e., $\indx_2(G)=3$, and this is achieved
by the following linear mappings. The broadcaster transmit $Y_1= X_2+X_3, Y_2= X_4+X_5$ and $Y_3= X_1+X_2+X_3+X_4+X_5.$
The decoding functions are, $X_1 = Y_1 + Y_2+Y_3; X_2 = Y_1+X_3; X_3 = Y_1+X_2; X_4 = Y_2+X_5; X_5 = Y_2+X_4.$

Although in general $\rdss_q(G) \ne n - \indx_q(G)$, these two quantities are
not too far from each other. In particular, for large enough alphabet, the left and right hand
sides can be arbitrarily close. This is reflected in Thm.~\ref{thm:main} below.

\subsection{Implication of the results of \cite{alon2008broadcasting}}
At this point we 
cast a result of \cite{alon2008broadcasting} in our context.
In \cite{alon2008broadcasting}, the problem of index coding was considered and 
to characterize the optimal size of an index code, 
the notion of a {\em confusion graph} was introduced. Two input strings, $\bfx= (x_1, \dots, x_n), \bfy = (y_1, \dots, y_n) \in \ff_q^n$
 are called {\em confusable} if there exists some $i \in \{1, \dots, n\}$, such that $x_i \ne y_i$, but 
 $x_j = y_j,$ for all $j \in N(i)$. In the confusion graph of $G$, the total number of vertices are $q^n$, and each vertex
 represents a different $q$-ary-string of length $n$. There exists an edge between two vertices
 if and only if the corresponding two strings are confusable with respect to the graph $G$.
 The maximum size of an {\em independent set} of the confusion graph is denoted by $\gamma(G)$.
 
 The confusion graph and $\gamma(G)$ in \cite{alon2008broadcasting} were used
 as  auxilaries to characterize the the {\em rate} of index coding; they were not used to model any 
 practical problem. 
 From our definition of  RDSS codes (see Def.~\ref{def:rdss}), it is evident that
this notion of {\em confusable} strings fits perfectly
 to the situation of {\em local recovery} of a distributed storage system. Namely, $\gamma(G)$, in our problem
 becomes the largest possible size of an RDSS code for a system with storage-graph given by $G$.

We restate one of the main theorems of \cite{alon2008broadcasting}
using the terminology we have introduced so far.
\begin{theorem}\label{thm:main}
Given a graph $G(V,E)$, we must have,
\begin{align}
n - &\rdss_q(G) \le \indx_q(G) \le n -\rdss_q(G)\nonumber \\
&  + \log_q\Big(\min\{n\ln q, 1+ \rdss_q(G)\ln q \}\Big). 
\end{align}
\end{theorem}
This result is purely graph-theoretic, the way it was presented
in \cite{alon2008broadcasting}. In particular, the size of maximum independent set
of the confusion graph, $\gamma(G)$ 
can be identified as the size of the RDSS code, and its relation to 
the {\em chromatic number} of the confusion graph, which represents the size of the
index code, was found.
 Namely the proof was dependent on the following two crucial
steps.
\begin{enumerate}
\item The {\em chromatic number} of the graph can only be so much away from
the {\em fractional chromatic number} (see, \cite{alon2008broadcasting} for detailed definition).
\item The confusion graph is {\em vertex transitive}. This implies that the maximum size of an 
independent set is equal to the number of vertices divided by the fractional chromatic number. 
\end{enumerate}
A proof of the first fact above can be found in \cite{lovasz1975ratio}.
In what follows, we give a simple {\em coding theoretic} proof of Thm.~\ref{thm:main}, where the technique is same as \cite{alon2008broadcasting}; but it   bypasses the graph-theoretic notations.  However, our proof will expose
some further nuances in the relation of index coding and RDSS codes (see, Sec.~\ref{sec:refine} and Lemma.~\ref{lem:linear}). Because of the
derandomization of  Lemma.~\ref{lem:linear}, we can get rid of a look-up table to decode the index code that is `dual' of a given RDSS code.


\subsection{The proof of the duality}\label{sec:dual}
We prove Theorem \ref{thm:main} with the help of following two lemmas. 
The first of them is immediate and can be proved by a simple averaging argument.
\begin{lemma}\label{lem:equiv}
If there exists an index code $\cC$ of length $\ell$ for a side information graph $G$ on $n$ vertices, then there exists an RDSS code
of dimension at least $n-\ell$ for the distributed storage graph $G$.
\end{lemma}
\begin{IEEEproof}
Suppose, the encoding and decoding functions of the index code $\cC$ are $f: \ff_q^n \to \ff_q^{\ell}$ and $g_i:\ff_q^{\ell+N(i)}\to \ff_q, i =1, \dots,n$.
There must exists some $\bfx \in \ff_q^\ell$ such that $|\{\bfy \in \ff_q^n: f(\bfy) = \bfx\}| \ge q^{n-\ell}$.
Let, $\cD_\bfx \equiv \{\bfy \in \ff_q^n: f(\bfy) = \bfx\}$ be the RDSS with
recovery functions,
$$
f_i(\{X_j, j \in N(i)\}) \equiv g_i(\bfx, \{X_j, j \in N(i)\}).
$$ 
\end{IEEEproof}
The second lemma might be of more interest as it is a bit less obvious.
\begin{lemma}\label{lem:rdss}
If there exists an RDSS code $\cC$ of dimension $k$ for a  distributed storage graph $G$ on $n$ vertices, then there exists an index code
of length $n-k +\log_q \min\{n\ln q, 1+ k\ln q \}$ for the side information graph $G$.
\end{lemma}
Combining these two lemmas we get the proof of Theorem \ref{thm:main} immediately.

To prove Lemma \ref{lem:rdss} , we need the help of two  other lemmas. 
First of all notice that, translation of any RDSS code is an RDSS code.
\begin{lemma}\label{lem:trans}
Suppose, $\cC\subseteq \ff_q^n$ is an RDSS code. Then any known translation of $\cC$ is also an 
RDSS code of same dimension. That is, for any $\bfa \in \ff_q^n$, $\cC+\bfa \equiv \{\bfy+\bfa: \bfy \in \cC\}$
is an RDSS code of dimension $\log_q |\cC|$.
\end{lemma}
\begin{IEEEproof}
Let, $(X_1, \dots, X_n) \in \cC.$ Also assume, $\bfa = (a_1, \dots, a_n)$, and $X'_i = X_i +a_i$.
 We know that, there
exist recovery functions such that,
$X_i = f_i(\{X_j: j \in N(i)\}).$
Now, $X'_i = X_i+a_i = f_i(\{X_j: j \in N(i)\}) + a_i \equiv f'_i(\{X'_j: j \in N(i)\}$.
\end{IEEEproof}

The proof of  Lemma \ref{lem:rdss}
crucially use the existence of a {\em covering} of the entire $\ff_q^n$, by translations of
an RDSS code. 
Indeed, we have the following result.
\begin{lemma}\label{lem:covering}
Suppose, $\cC\in \ff_q^n$ is an RDSS code for a graph $G$. There exists 
$m$ vectors $\bfx_j\in \ff_q^n, j =1, \dots, m$, such that
$$
\cup_{i =1}^{m} (\cC+\bfx_i )  = \ff_q^n
$$
where
$$
m =\frac{q^{n}}{|\cC|}\min\{n \ln q, 1+\ln |\cC|\}.
$$
\end{lemma}
\begin{IEEEproof}
Suppose, $\bfx_i, i =1, \dots, m$ are randomly and independently chosen from $\ff_q^n$.
Now, the expected number of points in the space not covered by any of the translations is at most
$
 q^n (1-|\cC|/q^n)^{m'} < q^n e^{-m'|\cC|/q^n} \le 1,
$
when we set $m' = q^{n-k} n \ln q \le m$ in the above expression (see \cite[Prop.~3.12]{babai1995automorphism}). 

If, instead we set $m' = \frac{q^{n}}{|\cC|}  \ln |\cC|$ then  the expected number of points, that do not belong to any of the $m'$ translations is at most $ \frac{q^{n}}{|\cC|} $.
To cover all these remaining points we  need at most $\frac{q^{n}}{|\cC|} $ other translations. Hence,
there must exists a covering such that $ \frac{q^{n}}{|\cC|}  \ln |\cC| + \frac{q^{n}}{|\cC|} = \frac{q^{n}}{|\cC|}(\ln |\cC| +1)\le m$ 
translations suffice.
\end{IEEEproof}
Using Lemmas \ref{lem:trans} and \ref{lem:covering} we now prove Lemma \ref{lem:rdss}.

\begin{IEEEproof}[Proof of Lemma \ref{lem:rdss}]
Lemmas \ref{lem:trans} and \ref{lem:covering} show that there exist, $\cC_1, \dots, \cC_{m},$ $\cC_i \subseteq \ff_q^n, i =1, \dots,m,$
all of which are RDSS codes of dimension $k$ such that
\begin{equation}\label{eq:cover}
\cup_{i =1}^{m} \cC_i  = \ff_q^n,
\end{equation}
where $m =  q^{n-k}\min\{n \ln q, 1+k\ln q\}$. Indeed, $\cC_i$ can set to be equal to $\cC+\bfx_i$ , which is an RDSS code by
 Lemma \ref{lem:trans}.

Now, any $\bfy \in \ff_q^n$ must belong to 
at least one of the $C_i$s. Suppose,  $\bfy \equiv (Y_1, \dots, Y_n)\in \ff_q^n$ and $\bfy \in C_i$. Then,
the encoding function of the desired index code $\cD$ is simply given by, $f(\bfy) = i$. 
 If the recovery functions of $\cC_i$ are $f^i_j, j=1, \dots, n$, then,   the decoding functions
of $\cD$ are given by:
$$
g_j(i, \{Y_l: l \in N(j)\}) =  f^i_j(\{Y_l: l \in N(j)\}).
$$
Clearly the length of the index code
is $\log_q m = n -k + \log_q (\min\{n\ln q,1+k\ln q\})$.
\end{IEEEproof}

The most crucial step in the proof of Thm.~\ref{thm:main} is Lemma~\ref{lem:covering}, that show existence
of a desired set of points in $\ff_q^n$:  we need to show
 the existence of a {\em covering} of the entire $\ff_q^n$, by translations of
an RDSS code. Next we show that stronger statements in lieu of Lemma~\ref{lem:covering}
is possible: the translations themselves form a
linear subspace. This leads to a derandomization and ease of decoding of the index code in each of the receivers.


\subsection{Refinements of Lemma \ref{lem:covering} and decoding of index code}\label{sec:refine}

In this section, we show that the $m$ points whose existence is guaranteed by Lemma \ref{lem:covering}
can be made to satisfy some extra properties. 
In particular, when $q=2$, any randomly chosen {\em linear} subspace of dimension $\log_2 m$
suffices for our purpose with high probability.
\begin{definition}
Given a set of vectors $\bfx_1, \dots, \bfx_\ell$ from $\ff_q^n$, define the {\em binary span} of 
the set to be $\{\sum_{i=1}^\ell a_i \bfx_i: (a_1,\dots, a_\ell) \in \{0,1\}^\ell \}.$
\end{definition}
\begin{lemma}\label{lem:linear}
Suppose, $\cC\subseteq \ff_q^n$ is an RDSS code for a graph $G$. There exists a set 
of $\ell = \log_2\frac{q^{n}}{|\cC|} +\log_2 (\min\{n \ln q, \ln (e|\cC|)\}) = \log_2\frac{q^{n}}{|\cC|}+ O(\log n)$ vectors, whose binary span
$\cD\subseteq \ff_q^n$ is 
such that
\begin{equation}\label{eq:covering}
\cup_{\bfx \in \cD} (\cC+\bfx )  = \ff_q^n.
\end{equation}
\end{lemma}
To  prove this lemma we construct a greedy algorithm that
chooses about $\log_2 m$ vectors  recursively  instead of $m$ random vectors of Lemma \ref{lem:covering}. 
The proof is deferred to the appendix.
The greedy covering argument that we employ in the proof was used to show the existence of good linear covering codes
in \cite{delsarte1986most} (see, also, \cite{goblick1963coding,cohen1983nonconstructive,mazumdar2010linear}).
We can use Lemma \ref{lem:linear} instead of Lemma \ref{lem:covering} to complete the proof of Lemma  \ref{lem:rdss}.
Lemma \ref{lem:linear} gives some algorithmic advantage
in decoding an index code that we explain next.

Suppose $\cC$ is an RDSS code with  known recovery functions. 
 Let $\cD$ be the set of vectors promised in Lemma \ref{lem:covering}
 such that $\cup_{\bfx \in \cD} (\cC+\bfx) = \ff_q^n$. Consider next the corresponding index code
 constructed in the proof of Lemma \ref{lem:rdss}. 
Given any $\bfy\in \ff_q^n$ as input,  the encoding of this index code finds a $\bfz \in \cD$ such that $\bfy \in \cC+\bfz.$
A bijection $\psi: \cD \to \ff_q^l$ that maps
 $\bfz$  to a $q$-ary vector of length $l = \log_q|\cD|$ completes the encoding of the index coding (here $l$ is the length of the index code\footnote{We assume $l$ to be an integer, which not necessarily is the case. The argument remains the same when $l$ is not an integer, except for the fact that we have to deal with ceiling and floor functions.}). In short, the encoding of the index code maps $\bfy$ to $\psi(\bfz)$ for an $\bfz: \bfy \in \cD+\bfz.$ 
Now for decoding of this index code, one first needs to map back any given  encoded vector $\bfu \in \ff_q^l$ to $\bfx' \equiv \psi^{-1}(u) \in \cD$, and
then use the recovery functions of the RDSS code $\cC+\bfx'$. The recovery functions of RDSS code $\cC+\bfx'$ is known, as they are known  for
the RDSS code $\cC$. 

In the above decoding of index code,  we must maintain a look-up table
of size exponential in $n$, that stores the bijective map $\psi^{-1}$ between $\ff_q^l$ to $\cD$. This map 
tells us recovery function of which RDSS code to use (among all the translations).
 However, using Lemma~\ref{lem:linear}
this constraint can be removed. 

Assume $g: \ff_q^{\log_q |\cD|} \to \ff_2^{\ell}$ is an arbitrary polynomial time bijective mapping that produces a binary sequence from a
$q$-ary sequence. There are many such mappings that can be trivially constructed.
 Using Lemma~\ref{lem:linear}, $\cD$ is the 
binary span of $\ell = \log_2|\cD|$ vectors $\{\bfd_1, \dots, \bfd_\ell\}$ such that $\cup_{\bfx \in \cD} (\cC+\bfx) = \ff_q^n$. Then the decoding of 
the obtained index code can be performed from $\bfu \in \ff_q^{\log_q |\cD|}$ in two steps. 
First, suppose $g(\bfu) = (a_1,\dots, a_\ell)$.
Next, we  compute $\bfx' = \sum_{i=1}^\ell a_i \bfd_i$. 
For the decoding of the index code, we now use the recovery functions of $\cC+\bfx'$.
The map from $\bfu$ to $\bfx'$ defines $\psi^{-1}$ in this case.
Hence, we no longer need to maintain a look-up table, and the required RDSS code, that we need to decode, 
can be found in polynomial time.


\begin{remark}
Note that, a random subset of $\ff_q^n$, generated as the binary span of $\log_2 m $ random and uniformly chosen vectors from $\ff_q^m$,
satisfies Eq.~\eqref{eq:covering} with high probability. 
This can be proved along the line of \cite{blin1990,cohen1997covering} where it was shown almost all linear codes 
are good covering codes.
\end{remark}

Given an RDSS code, our derandomization benefits only the decoding of the obtained  index code, and not the encoding. But also notice
that, encoding is performed by the broadcaster in one place, while decoding is performed in every receiver (that is likely to have less computational power
compared to the broadcaster).
 
 \section{Algorithmic results and constructions of RDSS codes}
 In this section we provide some constructions of RDSS codes, both for
 directed and undirected graphs. First, note that, existential results similar to Gilbert-Varshamov bound for
 codes can be provided for RDSS codes. 
 \begin{theorem}
For  the graph $G(V=\{1,\dots,n\},E)$, define,
$$
Q_q(G) = \{\bfx\in\ff_q^n : \exists i \text{ with } x_i \ne 0, x_j =0 \forall j \in N(i) \}.
$$
Then,
$$
\rdss_q(G) \ge n - \log_q(|Q_q(G)| +1).
$$
%
%
%
\end{theorem}
\begin{IEEEproof}
Recall that, any RDSS code can be found as an independent set of the confusion graph. The confusion graph  is regular
with degree exactly equal to $|Q_q(G)|$. Indeed, if for $\bfx,\bfy \in \ff_q^n$, $\bfy =\bfx+\bfv$ for some $\bfv \in Q_q(G)$, then 
$\bfx$ and $\bfy$ both cannot be part of an RDSS code without violating the repair condition.
Now, using Tur{\'a}n's Theorem, there must exist an RDSS code of size
$$
\frac{q^n}{|Q_q(G)| +1}.
$$
\end{IEEEproof}
$|Q_q(G)|$ can be bounded from above  in a number of  ways if some properties of the graph is known.
We give an example next.
\begin{example}[Degree distribution]
Using a simple union bound for counting, we get the following:
$$
|Q_q(G)| \le q^n(q-1) \sum_{i=1}^n \delta_i q^{-(i+1)},
$$
where $\delta_i$ is
the number of vertices with degree $i$. This shows that, the capacity of $G$ is at least,
$$
\rdss_q(G) \ge - \log_q [ (q-1)\sum_{i=1}^n \delta_i q^{-(i+1)}].
$$
For a large class of networks such as the internet, world-wide-web and social networks, the empirical 
degree distributions $\delta_i$ have been estimated (most of the times it follows a power-law decay).
Using these, the achievable storage-capacity of the networks can be approximated.
\end{example}
For general graphs, the union bound can be quite loose and it might be  difficult
to compute $|Q_q(G)|$. 
However, it is possible to construct codes and compute 
$\rdss_q(G)$ via deterministic algorithms using more sophisticated ways than above. 
We consider the cases of undirected and directed graphs separately as different algorithms are
needed in these scenarios. For impossibility results, however, the technique is same: we show that there
exists a large enough subset of vertices that cannot store any information on top of what the rest of the vertices 
already store.


 \subsection{Undirected graph}\label{sec:undirected}
In this section, we show that for an undirected graph $G$, an RDSS code can be constructed in polynomial time
that achieves a rate within half of what is optimal for $G$. In particular, if $G$ is bipartite, then 
the optimal code achieving a rate equal to $\rdss_q(G)$ can be constructed. Hence, for undirected graph 
it is relatively easy to compute or approximate $\rdss_q(G)$.

To achieve the above goal, start with the following lemma first. Recall that, a {\em vertex cover} of a graph $G(V,E)$ is
a subset $U\subseteq V$ such that $\forall (u,v) \in E$ either $u \in U$ or $v \in U$ or both.
\begin{lemma}\label{lem:vc}
For any undirected graph $G(V,E)$,  and any $q\ge 2$, 
\begin{equation}\label{eq:vc}
\rdss_q(G) \le VC(G),
\end{equation}
 where $VC(G)$ is the size of the minimum vertex cover of $G$.
\end{lemma}
\begin{IEEEproof}
Suppose, $A \subset V$ is an independent set in $G$. Any vertex $v \in A$ has $N(v) \subseteq V \setminus A.$ Hence,
$\rdss_q(G) \le n - |A|$. Notice, $V\setminus A$ is a vertex cover of $G$.  When $A$ is the largest independent set, we have,
   $\rdss_q(G) \le VC(G).$
\end{IEEEproof}

\subsubsection{Construction of code}
 A {\em matching} in a graph $G(V,E)$ is a set of edges such that no two edges share a common vertex.
The size of the largest possible matching of the graph $G$ is denoted by $M(G)$ below. Polynomial time algorithms to 
find the maximum matching is well-known \cite{edmonds1987paths}.

To store information in the graph, first we find a maximum matching $F \subset E$. Then for  any $(u,v)\in F,  u, v \in V$, we store the same 
variable in both $u$ and $v$. In this way we will be able to store $M(G)$ amount of information. Whenever one vertex fails we can go to only one other vertex to retrieve 
the information. Hence, $M(G) \le \rdss_q(G)$.

Surprisingly, this simple constructive scheme is optimum for bipartite graphs, within a factor 2 of optimum storage for arbitrary graphs
and is very unlikely to get improved upon via any other constructive scheme.

First of all, we need the following well-known lemma \cite{vazirani2001approximation}.
\begin{lemma}\label{lem:mat}
For any graph $G$, 
$$
M(G) \le VC(G) \le 2M(G).
$$
\end{lemma}
The proof is straight-forward. To cover all the edges one must include at least one vertex from the edges of any matching.
On the other hand, if both the endpoints of the edges of a maximal matching is deleted, no two other vertices 
can be connected (from the maximality of the matching). 

Now using Lemmas \ref{lem:vc}, \ref{lem:mat}, and the discussion above, we have,
\begin{equation}
M(G) \le  \rdss_q(G)\le VC(G) \le 2M(G).
\end{equation}
Hence,  for any  graph $G$, we can store via a constructive procedure $M(G) \ge \frac12 \rdss_q(G)$
amount of information. Indeed, for a $2$-approximation, we do not even need to find the maximum matching; 
a maximal matching, that can be found by a simple greedy algorithm, is sufficient.

It is unlikely that anything strictly better than the matching-code above can be found for  
an arbitrary graph $G$ in polynomial-time, because that would imply a better-than-2-approximation for the
minimum vertex cover. Khot and Regev \cite{khot2008vertex} have shown that if the unique game conjecture is
true then such algorithm is not possible. Inapproximability of minimum vertex cover under milder assumptions appear in the famous paper
of Dinur and Safra \cite{dinur2005hardness}.  


However for some particular classes of graphs we can do much better. Specifically if the graph $G$ is bipartite then 
K{\"o}nig's theorem asserts $M(G) = VC(G)$.
Hence for a bipartite graph $G$, $\rdss_q(G)= M(G)$ and an RDSS code can be designed in polynomial time. 

Other special graphs, such as planar graphs \cite{bar1982approximating,baker1994approximation}, that have better approximation algorithms 
for minimum vertex cover, might also allow us to approximate $\rdss_q(G)$ better. We left that exercise as future work.

\subsection{Directed graphs}\label{sec:directed}
Next we attempt to extend the above techniques to construct RDSS codes for directed graphs. 
The following proposition is a simple result that proves to be an useful converse bound.
\begin{proposition}\label{prop:fvs}
For any graph $G(V,E)$, and any $q\ge 2$,
\begin{equation}\label{eq:fvs}
\rdss_q(G) \le {\rm FVS}(G), 
\end{equation}
where ${\rm FVS}(G)$ is the minimum number of vertices to be removed to make $G$ acyclic (also called the minimum {\em feedback vertex 
set}.
\end{proposition}
Note that, results of \cite{bar2006index} or \cite{chaudhry2011complementary} imply that for any directed graph $G$,
$\indx_q(G)$ is at least the size of the maximum acyclic induced subgraph of $G$. From this, and from Thm.~\ref{thm:main}, we can deduce that
$\rdss_q(G) \le {\rm FVS}(G) + O(\log n)$. The above proposition is  stronger in the sense that we get rid of the
$\log$ term. 
\begin{IEEEproof}[Proof of Prop.~\ref{prop:fvs}]
Suppose, $U \subset V$ is such that the subgraph induced by $U$ is acyclic. We first claim that, the dimension of
any RDSS code in $G$  must be at most $|V\setminus U|$. 
Let us prove this claim with  a simple reasoning that appear in \cite{barg2013family}. Suppose $u \in U$ is such that all edges in $E$ that are outgoing from $u$ has the other end 
in $V \setminus U$. As the induced subgraph from $U$ is acyclic, there will always exist such vertex. Hence, whatever we store
in $u$, must be a function of what are stored in  vertices of $V\setminus U$.
Now, consider the subgraph induced by $U \setminus\{u\}$. As this subgraph is also acyclic, there must exist a vertex whose content is
a function of the the contents of   vertices of $V\setminus U$. Proceeding as this, we deduce that, no more than $|V\setminus U|$ amount
of information can be stored in the graph $G$.

Now consider  the maximum induced acyclic subgraph of $G$. If the vertex set of such subgraph is $U$, then $|V\setminus U| = {\rm FVS}(G).$
Hence, $\rdss_q(G) \le {\rm FVS}(G)$.
\end{IEEEproof}

It is not possible to construct a code by a matching, as in the case of undirected graph. In the undirected graph 
we could do that because, if $(u,v) \in E$, then just by replicating the symbol of $u$ in $v$ we can 
guarantee recovery for both $u$ and $v$. In the case of directed graph, such recovery is possible, if we have a
{\em directed cycle}: $u_0 \rightarrow u_1\rightarrow \dots \rightarrow u_{\ell-1} \rightarrow u_0$, where $u_i \in V$ and $(u_i,u_{(i+1)\mod \ell}) \in E$ for all $0\le i< \ell$.
 We can just store one symbol in $u_1,$
and then replicate this symbol over all vertices of the cycle. Whenever one node fails we can go to the next node in the cycle to recover what we lost.

Two cycles in the graph $G(V,E)$ will be called {\em vertex-disjoint} if they do not have a common vertex.
 
Suppose, $P$ is a set of vertex-disjoint cycles of the graph $G$. Then it is possible to store $|P|$ symbols
in the graph. Hence 
\begin{equation}
\rdss_q(G) \ge VD(G),
\end{equation}
 where $VD(G)$ is the maximum number of vertex-disjoint cycles
in the graph $G$.

At this point it would be  helpful to establish a relation between $VD(G)$ and $FVS(G)$. Such relation 
appear in the work of Erd{\"o}s and P{\'o}sa \cite{erdHos1965independent}. Namely, for any undirected graph, it was shown that
$
{\rm FVS}(G) \le VD(G) \log VD(G).
$
There are two bottlenecks of using this result for our purpose. First, this only holds for undirected graphs. Second,
computing the optimal vertex-disjoint cycle packing is a computationally hard problem even for undirected graphs.

There are a number of efforts towards generalizing the Erd{\"o}s and P{\'o}sa theorem for directed graphs culminating in 
\cite{reed1996packing} that shows that for directed graph there exists an increasing function $h:\integers \to \integers$ such that,
$$
{\rm FVS}(G) \le h(VD(G)).
$$
However, the function $h$ implied in \cite{reed1996packing} can be super-exponential. Hence, for our purpose it
is not of much interest.

In what follows, we show that a {\em fractional vertex-disjoint cover} also lead to an RDSS code. Albeit the code is {\em vector-linear}
as opposed to the scalar codes we have been considering so far. 
We need the following fractional vertex-disjoint packing result of Seymour  \cite{seymour1995packing}.
Suppose, $\cP$ is the set of all directed cycles of $G(V,E)$. Suppose, $\phi: \cP \to \rationals$ assigns a rational number to 
every directed cycle. Let $V(C), C \in \cP$ denote the vertices of the cycle $C$. We impose a condition that $\phi$ must satisfy,
$$
\sum_{C: v \in V(C)} \phi(C) \le 1,  
$$
for all $v \in G$. Under this condition we maximize the value of $\sum_{C\in \cP} \phi(C)$ over all functions $\phi$. Suppose this value is
$K$. Then \cite{seymour1995packing} asserts,
$$
FVS(G) \le 4K \ln 4K \ln\log_2 4K.
$$
We will now show a construction of RDSS codes using Seymour's result.
\begin{theorem}\label{thm:vector_con}
Suppose in each vertex of the directed graph $G(V,E)$ it is possible to store a vector of length $p$, i.e., from $\ff_q^p$, for a large enough integer 
$p$. Then, for any $q\ge2$,  it is possible to store constructively $pK$ $q$-ary symbols in the graph, such that content of any vertex can be recovered from its neighbors,
and
$$
4K \ln 4K \ln\log_2 4K \ge FVS(G) \ge \rdss_q(G).
$$ 
\end{theorem}
\begin{remark}
We can use the method  of \cite{chaudhry2011complementary}, where the {\em complementary index coding problem},
i.e., maximization of $n - \indx_q(G)$ is studied, to prove this theorem. Perhaps their result cannot be used as a blackbox
as that would lead to 
an extra additive error term of $O(\log \rdss_q(G))$, due to the gap between  $n - \indx_q(G)$ and
$\rdss_q(G)$. By a direct analysis, we can avoid this error term. However, 
the analysis of \cite{chaudhry2011complementary}   is more complicated than the proof below.  To find a vertex-disjoint packing in polynomial time
 the authors of \cite{chaudhry2011complementary} first constructs a so-called vertex-split graph and converts the 
vertex disjoint packing in to a edge-disjoint packing problem and then converts it back. It also
 uses crucially a result of \cite{nutov2004packing} to find a fractional edge-disjoint packing.
Below we follow a much simpler path.
\end{remark}

\begin{IEEEproof}[Proof of Thm.~\ref{thm:vector_con}]
Suppose, $\cP$ is the set of all directed cycles of $G(V,E)$, and $\phi: \cP \to \rationals$ is a function such that 
\begin{enumerate}
\item $\sum_{C: v \in V(C)} \phi(C) \le 1$, for all $v \in G$.
\item $\rdss_q(G) \le 4K \ln 4K \ln\log_2 4K,$ where $K =  \sum_{C\in \cP} \phi(C)$.
\end{enumerate}
We know such function $\phi$ exists from \cite{seymour1995packing} and Prop.~\ref{prop:fvs}.
Without loss of generality, we can assume $\phi(C) = \frac{n(C)}{p}$ for all $C \in \cP$,  $n: \cP \to \integers_{+}\cup\{0\}$, and $p$ is a positive
integer.

Suppose we want to store a vector $\bfx \in \ff_q^{pK}.$
In each vertex we store a vector of length at most $p$, i.e., content of each vertex belong to $\ff_q^p$.
These vectors are decided in the following way.
We partition the coordinates of $\bfx$, that is $[1,2, \dots, pK]$, in to $|\cP|$ parts. 
Each cycle $C \in \cP$ is assigned $n(C)$ coordinates to it. We can do such partition, because  $\sum_{C\in \cP} n(C) = pK$.
For any $C\in \cP$,  the $n(C)$ coordinates assigned to $C$ are stored in $v$ for all $v \in V(C)$. Hence the length of the vector
need to be stored in $v\in V$ is $\sum_{C: v \in V(C)} n(C) \le p$ which is consistent with our assumption. 

Now if the content  of any vertex $v$  is needed to be restored, we can use the contents of the neighboring vertices. 
If $v \in V(C)$, then the $n(C)$ symbols stored in $v$ can be restored from the copy stored in the 
vertex $u$ where $(v,u)$ is an edge in $C$.
This holds true for all $C \in \cP$ such that $v \in V(C)$.

The  function $\phi$ can be found by solving a linear program: maximize  $\sum_{C\in \cP} \phi(C)$, subject to
$\sum_{C: v \in V(C)} \phi(C) \le 1$, for all $v \in G$.
 The number of variables 
in the linear program is equal to 
the number of cycles in the graph $G$. 
The dual problem is given by means of finding a function $\psi:V \to \rationals$ that minimizes
 $\sum_{v\in V} \psi(v)$ such that $\sum_{v \in V(C)}\psi(v)\ge 1$ for every directed cycle $C$.
Although the number of constraints in this dual linear program can be
exponentially large, there exists a separation oracle that can differentiate between
a feasible solution and an infeasible one. For example, given any   $\psi:V \to \rationals$,
one can just calculate the shortest weight cycle, $\min_{C \in \cP}\sum_{v \in V(C)} v$,  in polynomial time and check 
whether that is greater than $1$ or not. 
If such separation oracle exists,  then the dual  linear program can be solved in polynomial time \cite[p.~102]{vazirani2001approximation}-- and at the same time
a primal optimal solution can also be found (by using say, ellipsoid method).
 
%
Hence, it is possible to explicitly construct the above-mentioned vector RDSS code. 
\end{IEEEproof}

Subsequently, we consider multiple node failures in our storage model.

\section{Multiple failures}\label{sec:gen}
In this section, we describe two possible generalizations of the quantity $\rdss_q(G)$ that are consistent with 
the distributed storage literature and take care of the situation when more than one server-nodes
simultaneously fail.

\subsection{Collaborative Local Repair on Graphs}

The notion of {\em cooperative local repair} was introduced as a generalization
of the definition of local recovery in \cite{rawat2014cooperative}. In this definition, instead of one server failure,
provisions for multiple server failures are kept. Next we extend this notion to distributed storage on graphs.

Given a graph $G(V=\{1,\dots,n\},E)$, we use each vertex to store a $q$-ary symbol.
A code $\cC\subseteq \ff_q^n$ is called  cooperative $t$-RDSS code  if 
 for any set of connected vertices $U \subset V, |U| \le t$,
there exist deterministic functions $f^U_i, i \in U$ such that
 for any codeword $(X_1,\dots, X_n)\in \cC$,
$X_i = f^U_i(X_{\cup_{l \in U}N(l) \setminus U})$
for all $i \in U$.
This means that if any set of $t$ or less connected vertices fail, then 
one should be able to recover them from the neighbors of that set.

Note that, it is necessary in the definition to consider all sets of size less than $t$ as well, because the local recovery of any  set $U, |U|= t$
does not imply that all proper subsets of $U$ are locally recoverable (i.e., not all neighbors of $U$ are neighbors of a given vertex in $U$). 

The reason it is sufficient to consider connected sets for the definition is 
that two disconnected sets of vertices of total size $t$ are locally recoverable as
any set less than size $t$ is.

We below consider as example only the special case of $t=2$ for undirected graphs.
In this case, apart from being a usual  RDSS code, the code must also be able to deal with the case
when both vertices of an edge fail. Hence the construction based on {\em matching} of Sec.~\ref{sec:undirected}
will not work. Instead, for our first result, we will need the following definition.

A {\em $k$-path} in a graph is a set of vertices $v_1, v_2, \dots, v_k$ such that $(v_i,v_{i+1})$ is an edge in the graph for
all $1\le i\le k-1$. A subset $S$ of vertices, such that for any $k$-path $\{v_1, v_2, \dots, v_k\}$ of the graph at least
one of $v_i$s must belong to $S$, is called a {\em  $k$-path vertex cover} \cite{brevsar2011minimum}. 

\begin{proposition}\label{prop:path}
Suppose, given an undirected graph $G(V,E), |V| =n$, $S\subset V$ is the smallest 
$3$-path vertex cover.
Then 
 the dimension of any cooperative $2$-RDSS code is at most $|S|$.
\end{proposition}
\begin{IEEEproof}
Assume, $W \subset V$ is such that every vertex in the the induced subgraph of $W$ has degree $1$ or $0$. Such sets are called
{\em dissociation set} and the size of smallest dissociation set is called the {\em dissociation number} \cite{yannakakis1981node}.
From the definition of cooperative $2$-RDSS codes, content of any vertex of $W$ can be reconstructed from 
vertices outside of $W$.
Then the dimension of any cooperative $2$-RDSS code is at most $ n - |W|$. On the other hand,
$V \setminus W$ is such that  for 
any $u,v,w \in V$:  $(u,v), (v,w) \in E$, at least one of $u,v$ or $w$ is in $V \setminus W$.
\end{IEEEproof}

In other words,   the dimension of any cooperative $2$-RDSS code is at most $n$ minus the dissociation 
number.
It is possible to find all vertex-disjoint $3$-paths in a graph $G$ in polynomial time \cite{williams2009finding}. Note that the smallest  
$3$-path vertex cover
 must contain at least one vertex from any $3$-path.
This allows us to construct a cooperative $2$-RDSS code that has dimension at least one-third of what is 
optimal possible. Indeed, we just repeat the same variable in all three vertices of a $3$-path.

To generalize the above procedure beyond $2$ erasures becomes cumbersome and also leads to
substantial loss in the dimension of RDSS codes. Instead, in the following,  we consider the usual scenario where
a provision of recovery from catastrophic failures is included via minimum distance of the code. 


\subsection{Considerations for Minimum distance}
Inclusion of the {\em minimum distance} as a necessary parameter in a locally repairable code
  is the norm in distributed storage \cite{gopalan2012locality}. In this subsection, on the RDSS codes,
  we further impose the constraint of minimum distance between the codewords.
  Given a graph $G(V,E)$ an {\em RDSS code with distance $d$} is an RDSS code $\cC \subseteq \ff_q^{|V|}$ such that for any $\bfx,\bfy \in \cC$,
 the Hamming distance between them, $\dist(\bfx,\bfy) \ge d$.
 
 By abusing notations slightly, for any graph $G(V,E)$ and any $U \subset V$, define $N(U)$ to
 be the set of all vertices in $V \setminus U$ that has at least one (incoming) edge from $U$. We have the following proposition.
 \begin{theorem}\label{thm:bound}
 For any graph $G(V,E)$, suppose there exists an RDSS code with distance $d$ and dimension $k$. Then,
 \begin{equation}\label{eq:general}
 d \le |V| - k + 1 - \max_{U \in \cI(G): |N(U)| \le k-1} |U| ,
 \end{equation}
 where for an undirected graph $\cI(G)$ is the set of all independent sets of $G$ and for directed graphs
 $\cI(G)$ is the set of vertex-sets of all induced acyclic subgraphs of $G$.
 \end{theorem}
 When no local recovery property is required, the graph $G$ can be thought as a complete graph. 
 In that case, the above bound reduces to the well-known Singleton bound of
 coding theory. When no distance property is required (i.e., $d=1$), the bound reduces to
 \begin{equation}\label{eq:special}
 k \le |V| -  \max_{U \in \cI(G): |N(U)| \le k-1} |U| .
 \end{equation}
 We claim that this implies 
  Equations
 \eqref{eq:vc} or \eqref{eq:fvs} (for the cases of undirected and directed graphs respectively). Let us show this for the
 case of undirected graphs as the case of directed graph is analogous. 
Assume that \eqref{eq:special} is satisfied but $k > VC(G)$. However this means that for the largest independent set $U^\ast \subset V$,
$|N(U^\ast)| \le |V \setminus U^\ast| = VC(G) <k$. Hence, from  \eqref{eq:special}, we have $k \le |V| -|U^\ast| =VC(G)$, which is a contradiction.
Hence, $k \le VC(G)$.

  Finally, when the graph is regular with degree $r$, the bound of \eqref{eq:general} becomes 
 \eqref{eq:locality}, as an independent set (or acyclic induced subgraph) $U'$ of size $ \Big\lfloor\frac{k-1}{r}\Big\rfloor= \Big\lceil\frac{k}{r}\Big\rceil- 1$ is guaranteed to exist via
 Tur{\'a}n's theorem. Indeed, Tur{\'a}n's theorem guarantees existence of an independent set of size $\frac{|V|}{r+1}$. Hence $k \le |V| - \frac{|V|}{r+1} = \frac{|V|r}{r+1}$. We therefore have, $\frac{|V|}{r+1} \ge \frac{k}{r} > \Big\lfloor\frac{k-1}{r}\Big\rfloor$. This guarantees existence of the independent set $U'$. 
 Note that, $N(U') \le k-1$ as the graph has degree $r$.
 
\begin{IEEEproof}[Proof of Thm.~\ref{thm:bound}]
The proof follows a generalization of the proof of Eq.~\ref{eq:locality} from  
 \cite{barg2013family, cadambe2013upper}. Below we provide the proof for undirected graphs which extends straightforwardly to directed graphs.

Let $\cC \in \ff_q^n, n = |V|$ be an RDSS code with distance $d$ and dimension $k$ for the graph $G$.
For any $I \subseteq V$, let $\cC_I$ denote the restriction of codewords of $\cC$ to the vertices of $I$.
 
 Suppose, $U \subset V$ is the largest  independent set such that $|N(U)| \le k-1$. Let $R$ be the $k-1$ sized subset
 that is formed by the union of $N(U)$ and any arbitrary $k-1 - |N(U)|$ vertices.
 Hence,
 $$
 |\cC_{U \cup R}| \le q^{k-1},
 $$
 which imply $d$ must be at most $n -|U \cup R |$.
 On the other hand $|U \cup R | = |U| + k-1$.
 This proves the theorem.
\end{IEEEproof}

%
%
 
 


The bound of \eqref{eq:general} can be made to be dependent on per node storage, or the alphabet size $q$. Indeed,
we can have the following proposition.
\begin{proposition}\label{prop:alpha}
 For any $q$-ary RDSS code on $G(V,E)$ with distance $d$ and dimension $k$,
\begin{equation}
k \le \min_{U \in \cI(G)} |N(U)| + \log_q\cA_q(|V| - |U \cup N(U)|, d)  \, ,
\end{equation}
where, $\cA_q(n,d)$ is the maximum size of a $q$-ary error-correcting code
of length $n$ and distance $d$, and $\cI(G)$ is  defined in Thm.~\ref{thm:bound}. 
\end{proposition}

\begin{IEEEproof}
As before, let $\cC \in \ff_q^n, n = |V|$ be an RDSS code with distance $d$ and dimension $k$ for the graph $G$.
We have, for any $U\in \cI(G)$,
$$
\cC_{U \cup N(U)} \le q^{|N(U)|}.
$$
Hence, there must exist an $\bfx \in \ff_q^{|U \cup N(U)|}$, such that $\cD \triangleq |\{\bfy \in \cC: \bfy_{U \cup N(U)} = \bfx\}| \ge q^{k -|N(U)|} $. Since, $\cD$
is a code with length $|V| - |U \cup N(U)|$ and minimum distance $d$, we must have,
$$
k -|N(U)| \le \log_q\cA_q(|V| - |U \cup N(U)|, d).
$$
\end{IEEEproof}

A  very elegant construction of locally repairable codes appeared in  \cite{barg2013family} that achieves the 
bound of \eqref{eq:locality}. That construction can also be used for RDSS codes with distance. We outline it below.

For construction of index codes, the   {\em clique partition}  method is a well-known
heuristic  \cite{chaudhry2008efficient}. 
This method can be easily adopted for construction of RDSS codes. Given a graph $G(V, E)$
the set of vertices are partitioned into minimum number of subsets, such that the subgraph induces by
any subset is a complete subgraph (or in to minimum number of cliques). If the size of any one clique is
$t$, then it is possible to store $t-1$ symbols in the vertices of the clique, with recovery guarantee from neighbors
in case of single failure. In this way it is possible to construct an RDSS code with dimension $n -{\rm CL}(G)$, 
where ${\rm CL}(G)$ is the minimum number of cliques that partition $V$. The construction of Sec.~\ref{sec:undirected}
via {\em matching} is a special case of clique-partition. However, as a downside, the problem of minimum clique-partition is
NP-complete (while there exists polynomial-time algorithms for matching).

On the other hand, if a clique-partition of a graph is given, then it is possible to construct an RDSS code with distance $d$
for that graph.  
Suppose, $V = V_1 \sqcup \dots \sqcup V_\ell$ such that the induced subgraphs
on $V_1,V_2, \dots, V_\ell$ are all cliques. 
Suppose $r_{\rm min}\ge 2$ is the size of the smallest clique in the partition.
By   \cite{barg2013family}, it is possible to construct a locally repairable code with locality $r = r_{\rm min}-1$
and length  $n$. Such code is a RDSS code with distance $d$ for $G$.

This construction will be good if the sizes of the cliques in the partition $V_1,V_2, \dots, V_\ell$ do not vary. If there is
large discrepancy among the sizes then it is still be possible to use the methods of  \cite{barg2013family}. In particular,
a method of constructing locally repairable code with disjoint repair groups of different sizes have been proposed in 
 \cite[Thm.~5.3]{barg2013family}, that can be adapted straight-forwardly for this scenario.

\vspace{0.1in}

\emph{Acknowledgements:}
We thank A. Agarwal, A. G. Dimakis and  K. Shanmugam for
 useful references. We also thank B. Saha for pointing out the constructive 
 nature of Thm.~\ref{thm:vector_con}.

\bibliographystyle{abbrv}
\bibliography{aryabib}

\appendix
\subsection{Proof of Lemma \ref{lem:linear}}
We prove following general statement below which will imply Lemma \ref{lem:linear}.

\begin{proposition}\label{prop:cover}
For every subset 
$\cF \subseteq \ff_q^n$, there exists a 
 set $\cD \in \ff_q^n$  that  is binary span of $\log_2(q^n |\cF|^{-1}\min\{n \ln q, 1+\ln |\cF| \})$ vectors and
$$
\cup_{\bfx \in \cD} (\cF+\bfx) = \ff_q^n.
$$
\end{proposition}
The proof is contingent on the following result from classical coding theory.
\begin{lemma}[Bassalygo-Elias]\label{lem:elias}
Suppose, $\cC,\cB \subseteq  \ff_q^n.$ Then,
\begin{equation}
\sum_{\bfx \in \ff_q^n} \mid(\cC+\bfx) \cap \cB\mid = |\cC| |\cB|.
\end{equation}
\end{lemma}
\begin{IEEEproof}
\begin{align*}
\sum_{\bfx \in \ff_q^n} \mid(\cC+\bfx) \cap \cB\mid & = |\{(\bfx,\bfy): \bfx \in \ff_q^n, \bfy \in \cB, \bfy \in \cC+\bfx\} | \\
& = |\{(\bfx,\bfy): \bfx \in \ff_q^n, \bfy \in \cB, \bfx \in \bfy - \cC\} |\\
& = |\{(\bfx,\bfy):  \bfy \in \cB, \bfx \in \bfy - \cC\} |\\
& = |\cB| |\bfy - \cC|
 = |\cC| |\cB|,
\end{align*}
where $\bfy - \cC \equiv \{\bfy-\bfa: \bfa \in \cC\}$.
\end{IEEEproof}
Now, for any set $\cF \subseteq \ff_q^n$, define 
\begin{equation}\label{eq:q}
Q(\cF) \equiv 1 - \frac{|\cF|}{q^n}.
\end{equation}
In words, $Q(\cF)$ denote the proportion of $\ff_q^n$ that is not covered by $\cF$.
The following property is a result of Lemma \ref{lem:elias}.
\begin{lemma}\label{lem:q}
For every subset $\cF \subseteq \ff_q^n$, 
\begin{equation}
q^{-n}\sum_{\bfx\in \ff_q^n} Q(\cF\cup (\cF+\bfx)) = Q(\cF)^2.
\end{equation}
\end{lemma}
\begin{IEEEproof}
We have,
$$
|\cF\cup (\cF+\bfx)| = 2|\cF| - |\cF\cap (\cF+\bfx)|.
$$
Therefore,
$$
 Q(\cF\cup (\cF+\bfx)) = 1 - 2|\cF| q^{-n} +  |\cF\cap (\cF+\bfx)|q^{-n},
$$
and hence,
\begin{align*}
q^{-n}\sum_{\bfx\in \ff_q^n} Q(\cF\cup (\cF+\bfx)) &= 1 - 2|\cF| q^{-n} \\ &\quad+ q^{-2n}\sum_{\bfx\in \ff_q^n}|\cF\cap (\cF+\bfx)|\\
&= 1 - 2|\cF| q^{-n} + q^{-2n}|\cF|^2\\
& = (1 - |\cF| q^{-n})^2,
\end{align*}
where in the second line we have used Lemma \ref{lem:elias}.
\end{IEEEproof}

Now we are ready to proof Prop.~\ref{prop:cover}.

\begin{IEEEproof}[Proof of Prop.~\ref{prop:cover}]
From Lemma \ref{lem:q}, for every subset 
$\cF \subseteq \ff_q^n$, there exists $\bfx \in \ff_q^n$ such that
$$
Q(\cF\cup (\cF+\bfx)) \le  Q(\cF)^2.
$$
For the set $\cF\equiv \cF_0$, recursively define, for i =1, 2,\dots
$$
\cF_i = \cF_{i-1} \cup (\cF_{i-1} + \bfz_{i-1}),
$$
where $\bfz_i\in \ff_q^n$ is such that, 
$$
Q(\cF_i\cup (\cF_i+\bfz_i)) \le  Q(\cF_i)^2, \quad i =0,1,\dots
$$
Clearly,
$$
Q(\cF_t) \le Q(\cF_0)^{2^t} = \Big(1- q^{-n} |\cF|\Big)^{2^t} \le e^{- q^{-n} |\cF|2^t}.
$$
At this point we can just use the argument of the proof of Lemma \ref{lem:covering}, with $2^t$ playing the role of $m'$ therein.

\begin{figure}[htbp]
\begin{center}
\includegraphics[width=0.5\textwidth]{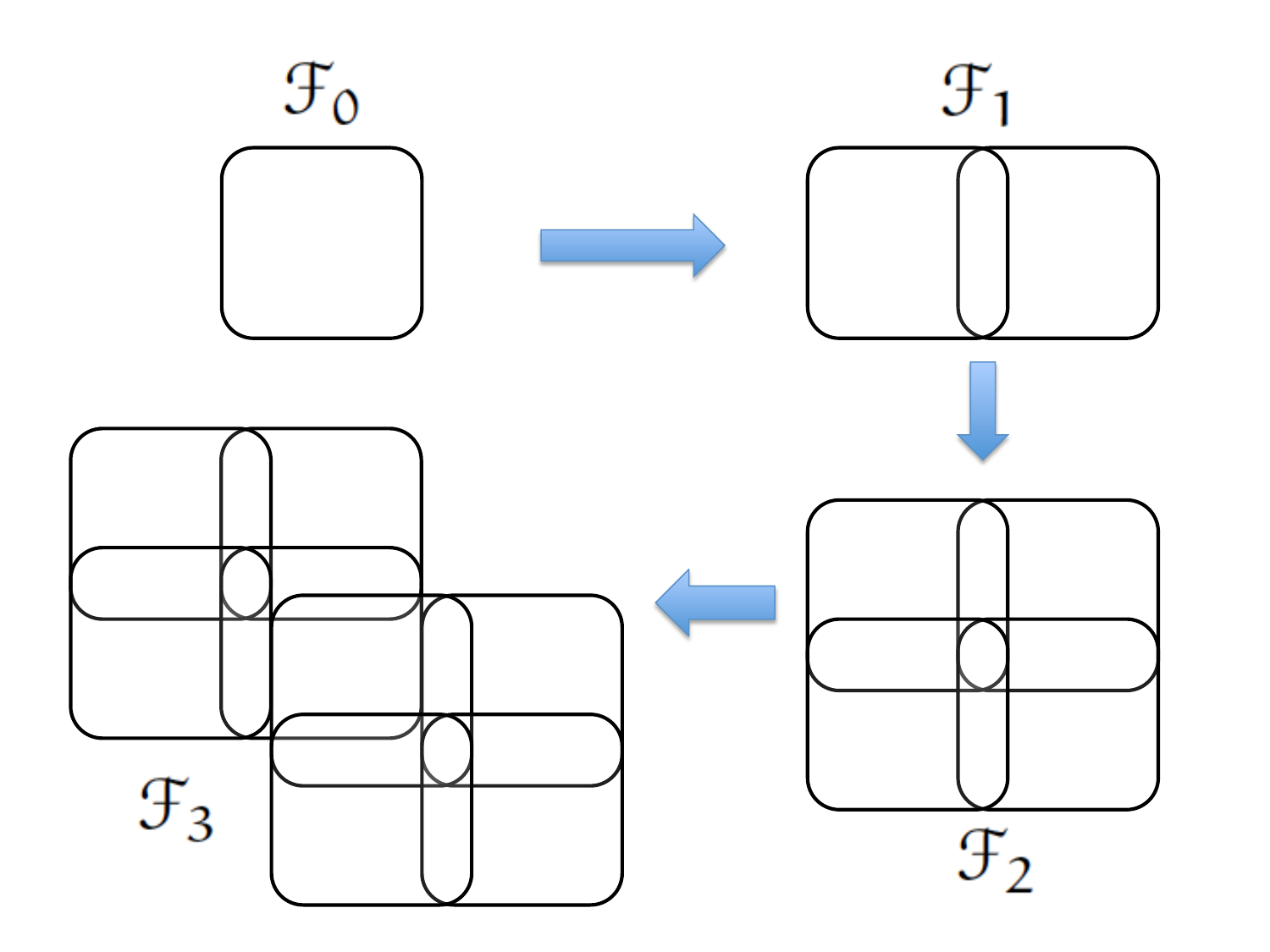}
\end{center}
\caption{The recursive construction of the sets $\cF_1, \cF_2, \cF_3$ of Prop.~\ref{prop:cover}.} 
\label{fig:replicate}
\end{figure}

On the other hand $\cF_t$ contains $\cF_0$ and its $2^t -1$ translations (see, Figure \ref{fig:replicate} for 
an illustration). 
Hence,  there exists $m = \min\Big\{\frac{q^nn \ln q}{|\cF|}, \frac{q^n (1+\ln |\cF|)}{|\cF|}\Big\}$ vectors
$\bfx_0 =0, \bfx_1, \bfx_2, \dots, \bfx_{m-1} \in \ff_q^n$, such that
$$
\cup_{i=0}^{m-1} (\cF+\bfx_i) = \ff_q^n.
$$
These $m$ vectors form the binary span of the $t$ vectors we have chosen via the greedy procedure. 
\end{IEEEproof}

\begin{IEEEbiographynophoto}{Arya Mazumdar} (S'05-M'13)
 is an assistant professor in University of Minnesota-Twin Cities (UMN). Before coming to UMN, he was a postdoctoral scholar
  at the Massachusetts Institute of Technology. He received the Ph.D. degree 
   from  University of Maryland, College Park, in 2011. 

Arya is a recipient of the NSF CAREER award, 2015 and  the 2010 IEEE ISIT Student Paper Award. He is also the recipient of the Distinguished Dissertation Fellowship Award, 2011, at the University of Maryland.    
   He spent the summers of 2008 and 2010 at the Hewlett-Packard Laboratories, Palo Alto, CA, and IBM Almaden Research Center, San Jose, CA, respectively.
 Arya's research interests include error-correcting codes, information theory and their applications.
 \end{IEEEbiographynophoto}

\balance
\end{document}